\documentclass{aa}
\usepackage{color}
\usepackage{graphicx}
\usepackage[colorlinks=true,linkcolor=blue, urlcolor=blue, citecolor=blue]{hyperref}
\usepackage{txfonts}
\usepackage[utf8]{inputenc}
\usepackage{amsmath}
\usepackage{amssymb}

\allowdisplaybreaks[4]
\begin{document} 

\title{The aperiodic firehose instability of counter-beaming electrons \\ in space plasmas}

	\titlerunning{Aperiodic firehose instability of beaming electrons}

\author{M. Lazar\inst{1,2} \fnmsep\thanks{\email{mlazar@tp4.rub.de}}, R.~A. L\'{o}pez \inst{1,3}, P.~S. Moya$^{1,4}$, S. Poedts$^{2,5}$, and S.~M. Shaaban\inst{6}} 
         \authorrunning{M.\ Lazar et al.}

\institute{
Centre for mathematical Plasma Astrophysics/Dept.\ of Mathematics, Celestijnenlaan 200B, 3001 Leuven, Belgium, \\ \email{marian.lazar@kuleuven.be}
\and
Institute for Theoretical Physics IV, Faculty for Physics and Astronomy, Ruhr-University Bochum, D-44780 Bochum, Germany
\and
Departamento de F\'{\i}sica, Universidad de Santiago de Chile, Usach, Santiago, Chile 
\and
Departamento de F\'{\i}sica, Facultad de Ciencias, Universidad de Chile, Santiago, Chile
\and
Institute of Physics, University of Maria Curie-Sk{\l}odowska, Pl.\ M.\ Curie-Sk{\l}odowska 5, 20-031 Lublin, Poland
\and
Theoretical Physics Research Group, Physics Department, Faculty of Science, Mansoura University, 35516, Mansoura, Egypt
}

   \date{\today}
   
  \abstract
   {Recent studies have revealed new unstable regimes of the counter-beaming electrons specific to hot and dilute plasmas from astrophysical scenarios: An aperiodic firehose-like instability is induced for highly oblique angles of propagation relative to the magnetic field, resembling the fast growing and aperiodic mode triggered by the temperature anisotropy $T_\parallel > T_\perp$ (where $\parallel, \perp$ denote directions relative to the magnetic field). 
   }
   {The counter-beaming electron firehose instability is investigated here for space plasma conditions, that includes not only a specific plasma parameterization but, in particular, the influence of an embedding background plasma of electrons and ions (protons).}
   {Fundamental plasma kinetic theory is applied to prescribe the unstable regimes, and characterize the wave-number dispersion of the growth rates, and differentiate from the regimes of interplay with other instabilities. We also use numerical particle-in-cell simulations to confirm the instability of these aperiodic modes, and their effects on the relaxation of counter-beaming electrons.}
   {Linear theory predicts a systematic inhibition of the (counter-)beaming electron firehose instability (BEFI), by reducing the growth rates and the range of unstable wave-number with increasing the relative density of the background electrons. To obtain finite and reasonably high values of the growth rate, the (relative) beam speed does not need to be very high (just comparable to the thermal speed), but the (counter-)beams must be dense enough, with a relative density at least 15-20\% of the total density. Quantified in terms of the beam speed and the beta parameter the plasma parametric conditions favorable to this instability are also markedly reduced under the influence of background electrons. Numerical simulations confirm not only that BEFI can be excited in the presence of background electrons, but also the inhibiting effect of this population, especially when this is cooler. In the regimes of transition to electrostatic (ES) instabilities, BEFI is still robust enough to develop as a secondary instability, after the relaxation of beams under a quick interaction with ES fluctuations.}
  {To the features presented in previous studies, we can add that BEFI resembles the properties of solar wind firehose heat-flux instability triggered along the magnetic field by the anti-sunward electron strahl. However, BEFI is driven by a double (counterbeaming) electron strahl, and develops at highly oblique angles, which makes it potentially effective in the regularization and relaxation of the electron counter-beams observed in expanding coronal loops (with closed magnetic field topology) and in interplanetary shocks.}

    \maketitle

\section{Introduction} \label{sec:intro}

The wave instabilities generated by electron beams in various plasma setups are invoked in many applications in astrophysics and space plasmas. Magneto-genesis in galaxies and intergalactic medium can be explained by the Weibel-type instabilities involving interpenetrating electron beams, which can produce the primordial magnetic field seeds \citep{Schlickeiser2005, Lazar2009} to be amplified by cosmological dynamos \citep{Beck-2015}. 
Nonthermal emissions from cosmological sources, e.g., gamma-ray bursts, active galactic nuclei, pulsars, etc., often invoke two-stream electrostatic (ES) or electromagnetic (EM) instabilities for the relaxation of the relativistic electron jets \citep{Stockem-etal-2007, Bret-2009, Schlickeiser-etal-2013}. 
Radio emissions (e.g., type-II or type-III bursts) associated to solar flares and heliospheric shocks driven by coronal mass ejections (CMEs), are believed to be the result of an ES Langmuir relaxation of electron beams with energy in the range of a few hundred keV \citep{Ganse-etal-2012, Lee-etal-2019}. 
Less energetic (up to a few keV) but more thermalized beams, known as strahls, are more recurrent in the solar wind, undergoing a continuous erosion with heliocentric distance \citep{Maksimovic-etal-2005, Anderson-etal-2012}, most probably due to the self-induced instabilities and wave fluctuations \citep{Verscharen-etal-2019, Che2019, Micera-etal-2020}, whose nature highly depend on the properties of electron strahls \citep{Lopez-etal-2020a}.

Recently, \cite{Lopez-etal-2020b} have revealed the existence of a firehose-like aperiodic instability driven by two symmetric populations of electrons, counter-moving along the regular magnetic field. It develops only at oblique angles of propagation with respect to the regular magnetic field, and remains aperiodic only if the electron counter-beams are perfectly symmetric, namely, with the same number densities, the same temperatures and the same beaming or drifting speeds. 
These properties and the fact that the trigger is an excess of kinetic (free) energy of electrons in parallel direction, makes this instability similar to the aperiodic firehose instability driven by a temperature anisotropy $T_\parallel > T_\perp$ with $\parallel, \perp$ denoting directions relative to the magnetic field) \citep{Li2000, Gary2003, Camporeale2008, Shaaban2019, Lopez2019a, Moya-etal-2022}. 
However, the ignition of this instability is conditioned by a relatively small drift ($v_d$) between counter-beaming electrons, that should not exceed the electron thermal speed ($\alpha_e$), otherwise (if, e.g., $v_d > \alpha_e$), it is highly competed by the ES instabilities. Indeed, recent results provided by \citet{Moya-etal-2022} have shown that in the $v_d/\alpha_e >1$ regime, quasi-parallel ES instabilities dominate with growth rates about one order of magnitude larger than the aperiodic (oblique) firehose instability \footnote{For a more fluid-like regime of electron (counter-)beams the instability wave spectra is dominated by the ES fluctuations (with the highest growth rates), while the EM or hybrid fluctuations may be predicted either for low drifts, i.e., $v_d < \alpha_e$, or for very high, relativistic beaming speeds corresponding to more than 100~keV \citep{Gary2003, Lazar2009, Bret-2009, Jao-Hau-2016}.}. 
%

Focusing on solar plasma outflows, the observations have revealed electron counter-beams, parallel and anti-parallel to the local magnetic field (in the literature also known as couterstreaming or bi-directional electrons, or even double strahls), since the first satellite missions. e.g., VELA~5 and 6, and IMP~6 \citep{Montgomery-etal-1974},  ISEE~3 \citep{Bame-etal-1981,Gosling-etal-1987}, but also by heliospheric spacecraft, such as Helios 1 and 2 \citep{Pilipp-etal-1987,Macneil-etal-2020}, as well as Ulysses outside the ecliptic \citep{Hammond-etal-1996,Lazar-etal-2014}.
Electron counter-beams have also been reported by the more recent missions, e.g., ACE \citep{Skoug-etal-2000,Steinberg-etal-2005,Anderson-etal-2012}, Wind \citep{Larson-etal-1996,Fitzenreiter-etal-2003, Nieves-Chinchilla-2008}, STEREO A/B \citep{Lavraud-etal-2010,Kajdic-etal-2013,Carcaboso-etal-2020}. 
In situ observations associate these electron counter-beams mainly with the foreshock or upstream regions of the bowshock \citep{Fitzenreiter-etal-2003} and interplanetary shocks driven by corotating interactions regions (CIRs) \citep{Lavraud-etal-2010} and coronal plasma ejection (CMEs) \citep{Lazar-etal-2014, Cremades-etal-2015}. More energetic beams are also associated with closed magnetic topology typical of CMEs with expanding coronal loops, more symmetric counter-beams at the apex, and less symmetric on flanks \citep{Skoug-etal-2000, Lazar-etal-2014}. 
The properties of electron (counter-)beams can be determined either directly from the in-situ measurements of the velocity distributions \citep{Maksimovic-etal-2005, Nieves-Chinchilla-2008, Bercic-etal-2019}, or indirectly, either from the analysis of the enhanced fluctuations detected also in-situ \citep{Pulupa-Bale-2008}, or from the analysis of radio (or even harder) emissions of different nature, e.g., type-II bursts from upstream of the shock, or type-IV bursts related to the CME magnetic clouds \citep{Pick-Vilmer-2008}. The radiative or non-radiative relaxation of electron (counter-)beams may involve a wide palette of plasma wave instabilities, which depend on the properties of these beams including, as already mentioned above, instabilities of ES, EM or hybrid waves, periodic or aperiodic, and propagating parallel or obliquely to the magnetic field \citep{Verscharen-etal-2019, Lopez-etal-2020a,Moya-etal-2022}.

Here we investigate the aperiodic firehose instability driven by the counter-beaming electrons \citep{Lopez-etal-2020b,Moya-etal-2022}, taking into account that under common conditions in the heliosphere, the electron counter-beams are embedded in a background plasma of stationary electrons and ions, mainly protons \citep{Fitzenreiter-etal-2003, Lavraud-etal-2010, Anderson-etal-2012, Carcaboso-etal-2020}. 
We will call it the beaming electron firehose instability, BEFI, for short.
In the next section we introduce the model for the electron distribution function, with two counter-beams and a stationary background population, as well as the analyzed parametric cases. The results from the linear kinetic theory of wave instabilities are presented in section 3. We analyze both the dispersion of the frequency and the growth rate, as a function of the wave number, but also the threshold conditions of BEFI as a function of the main parameters of the plasma. Moreover, Section 4 presents the results from the numerical simulations which, at this moment, seek to bring a confirmation of the predictions of the linear theory, and to qualitatively describe the temporal evolution of BEFI. 
In the last section, we conclude our results, and discuss potential implications of this instability in heliospheric applications.

\section{Plasma model and parameters} \label{model}

We consider a plasma of electrons (subscript $e$) and protons (subscript $p$), the dominant plasma species in the solar outflows. Of interest for present study are events revealing counterbeaming electrons, e.g., associated with interplanetary shocks, CIRs and CME closed magnetic fields, and where the electron velocity distributions (VDs) exhibit three distinct populations
\begin{align}
  f_e\left({ v_{\perp},v_{\parallel}}\right) = &
  \frac{n_{0}}{n_e}~f_{0}\left({ v_{\perp},v_{\parallel}}\right)
  +\frac{n_{1}}{n_e}~f_{1}\left({ v_{\perp},v_{\parallel}}\right) + \frac{n_{2}}{n_e}~f_{2}\left({ v_{\perp},v_{\parallel}}\right). \label{e1}
\end{align}
With numbering subscripts we indicate the stationary background component - subscript 0, and the more or less symmetric counterbeams - subscripts 1 and 2. Relative densities $n_j/n_e$ ($j=0, 1, 2$) are defined with respect to $n_e$, the total electron number density, equal to the proton density $n_e=n_p$.

For each individual beam the VD is assumed a temperature isotropic drifting Maxwellian of the form
\begin{equation}
f_j(v_\perp,v_\parallel)=\frac{1}{\pi^{3/2}\alpha_j^3}
\exp\left\{-\frac{v_\perp^2}{\alpha_j^2}
-\frac{(v_\parallel-U_j)^2}{\alpha_j^2}\right\}\,, \label{e2}
\end{equation}
where $\alpha_j=(2k_BT_j/m_e)^{1/2}$ is the thermal velocity and $U_j$ the drift velocity of the $j$-th beam. Using the zero-current condition, we find the drift velocities related by $n_2 U_2= -n_1 U_1$. 
Stationary or non-drifting ($U_0 = 0$) background electrons are modelled by an isotropic Maxwellian distribution, and can be cooler or hotter than electron beams. 
Protons are also considered a stationary (neutralizing) background, described by an isotropic non-drifting Maxwellian VD, and with the same temperature as the background electrons.

We also assume the electron counter-beams (subscript $b$ in the next) sufficiently symmetric, i.e., with the same relative drifts, $|U_1| = |U_2| = U_b$, the same number densities, $n_1 = n_2 = n_b = (n_e-n_0)/2$, and the same thermal velocities, $\alpha_1 = \alpha_2 = \alpha_b$. 
Although in CIRs and interplanetary shocks asymmetric counter-beams are far more likely to occur, the symmetry considered here allows us to reduce the parameters space, and thus focus on the effects of the background electron population, for a series of new cases obtained by varying the main properties of electron populations, which contrast to the previous results reported  by \cite{Lopez-etal-2020b} and \cite{Moya-etal-2022} for $n_0 = 0$. 
The dispersion and stability properties are investigated for different parametrizations of electron populations, established by keeping constant the relative beaming speed $U_b = 0.065 c$ (where $c$ is the speed of light in vacuum), and varying thermal velocities $\alpha_0$ and $\alpha_b$, and also relative number densities, e.g., $n_b/n_e$. 
Table~\ref{t1} shows the parameters for the most relevant plasma configurations that we have analyzed. These are classified in four cases corresponding to different thermal velocities $\alpha_0$ and $\alpha_b$ (in units of $c$), and for each of them a number of four subcases defined by different relative number densities of the beams $n_b/n_e$. 
It should be noted that we also chose thermal velocities slightly lower than the beam or drift velocity, regimes for which previous qualitative estimates indicated a possible competition with electrostatic instabilities (parallel to the magnetic field). Here we will also discuss these regimes of transition from the dominance of BEFI to that of electrostatic instabilities, which are found to be sensitive not only to the properties of the counter-beams, but also to the presence of background electrons.


First cases with a lower relative density of the electron counter-beams are more relevant to the space plasma conditions, including conditions at CIRs, CMEs and interplanetary shocks, while the other cases with a lower density of the background electrons are more close to the configuration studied by \cite{Lopez-etal-2020b}. 
For similar relative beaming speeds, with, e.g. $U_b/c = $ 0.06, 0.065, 0.07, and moderate values of plasma betas $\beta_e =$ 2, 4, but in the absence of the background population of electrons, linear theory predicts high enough maximum growth rates of beaming electron firehose instability (BEFI), and PIC simulations confirm that this unstable mode develops and can be faster than electrostatic instabilities \citep{Lopez-etal-2020b,Moya-etal-2022}.  
\begin{table}[t!]
\caption{\label{t1} Four cases corresponding to different thermal velocities $\alpha_0$ and $\alpha_b$ (in units of $c$), and for each case other four distinct sub-cases as defined by different relative number densities of the electron background $n_0/n_e$ or beams $n_b/n_e$. For all cases $U_b/c = 0.065$ and $\beta_p = 1.96$.}
\centering
\begin{tabular}{l c c c c c c}
\hline\hline
 Case &  $\alpha_0/c$ & $\alpha_b/c$ & $n_0/n_e$ & $n_b/n_e$ & $\beta_0$& $\beta_b$ \\ [0.5ex]
\hline
 1.a & 0.07 & 0.07 & 0.6  & 0.20 & 1.176 & 0.392 \\
 1.b & 0.07 & 0.07 & 0.5  & 0.25 & 0.980 & 0.490 \\
 1.c & 0.07 & 0.07 & 0.3 & 0.35 & 0.588 & 0.686 \\
 1.d & 0.07 & 0.07 & 0.1 & 0.45 & 0.196 & 0.882 \\
 \hline
 2.a & 0.04 & 0.07 & 0.6 & 0.20 & 0.384 & 0.392 \\
 2.b & 0.04 & 0.07 & 0.5 & 0.25 & 0.320 & 0.490 \\
 2.c & 0.04 & 0.07 & 0.4 & 0.30 & 0.256 & 0.588 \\
 2.d & 0.04 & 0.07 & 0.1 & 0.45 & 0.064 & 0.882 \\
  \hline
 3.a & 0.02 & 0.07 & 0.6 & 0.20 & 0.096 & 0.392 \\
 3.b & 0.02 & 0.07 & 0.5 & 0.25& 0.08  & 0.490 \\
 3.c & 0.02 & 0.07 & 0.3 & 0.35 & 0.048 & 0.686 \\
 3.d & 0.02 & 0.07 & 0.1 & 0.45 & 0.016 & 0.882 \\
  \hline
 4.a & 0.07 & 0.04 & 0.6 & 0.20 & 1.176 & 0.128 \\
 4.b & 0.07 & 0.04 & 0.5 & 0.25 & 0.980 & 0.160 \\
 4.c & 0.07 & 0.04 & 0.4 & 0.30 & 0.784 & 0.192 \\
 4.d & 0.07 & 0.04 & 0.3 & 0.35 & 0.588 & 0.224 \\
 \hline
\end{tabular} 
\end{table}

In the present analysis, the plasma beta parameter $\beta_j = 8 \pi n_j k_B T_j / B_0^2$ ($j= 0, b$) is calculated for each electron population using the corresponding number density $n_j$ and temperature $T_j$. 
Our parameterization, see Table~\ref{t1}, attempts to cover conditions specific to interplanetary shocks triggered by the fast winds, e.g., in CIRs, where $\beta_e \gtrsim 1$, but also the low $\beta_e \lesssim 1$ conditions, more characteristic to CMEs. 
Calculated with the total number density the plasma frequency $\omega_{p,e}^2 = (4 \pi n_e e^2/m_e)^{1/2}$ intervenes in the normalization of the wave number, while the electron gyrofrequency $\Omega_e = |e| B_0/(m_e c)$ in the normalization of wave frequency and growth rate. 
For the plasma frequency to gyrofrequency ratio we consider $\omega_{p,e}/\Omega_e = 20$, which is relevant for solar wind conditions, and ensures a reasonable computational time in the numerical simulations (see section \ref{pic}). We analyze the full spectrum of unstable modes, triggered by the relative drift of the counter-beams (for all angles of propagation with respect to the magnetic field), by using the dispersion solver developed in \cite{Lopez2019b} and \cite{Lopez-etal-2021}. 
Previous studies have shown that BEFI is mainly conditioned by the ratio between this drift, $U_b$, and the thermal velocity of the beaming electrons, $\alpha_b$. 
Thus, if $\alpha_b$ is higher, and, implicitly, the corresponding $\beta_b$ is higher, then $U_b$ must also increase in order to excite the instability \citep{Lopez-etal-2020b, Moya-etal-2022}. 
This condition shapes the drifting velocity thresholds of BEFI, see Fig.~3 in \cite{Lopez-etal-2020b}, and resembles that of the firehose heat-flux instability induced at  (quasi-)parallel angles of propagation by the uni-directional electron strahls/beams (carrying the main heat-flux in the solar wind), see, e.g., Fig.~11 in \cite{Shaaban2018}.

%
%
%

\section{Linear theory} \label{befi}

\subsection{BEFI with background electrons: cases 1 and 2}

We first discuss the linear dispersion properties of BEFI, through a parametric analysis that allows the characterization of different regimes of this instability, when predicted to be dominant or in competition with other unstable modes. 
Table~\ref{t1} presents plasma configurations which are found relevant for the existence of BEFI. These are classified into four cases, corresponding to different thermal velocities $\alpha_0$ and $\alpha_b$ (in units of $c$), and other four distinct sub-cases, as defined by various relative number densities of the electron background $n_0/n_e$ (or electron beams $n_b/n_e$). For all cases we consider the same relative drift $U_b/c = 0.065$, and for protons $\beta_p = 1.96$. 

\begin{figure*}[t!]
  \begin{center}
    \includegraphics[width=0.9\textwidth]{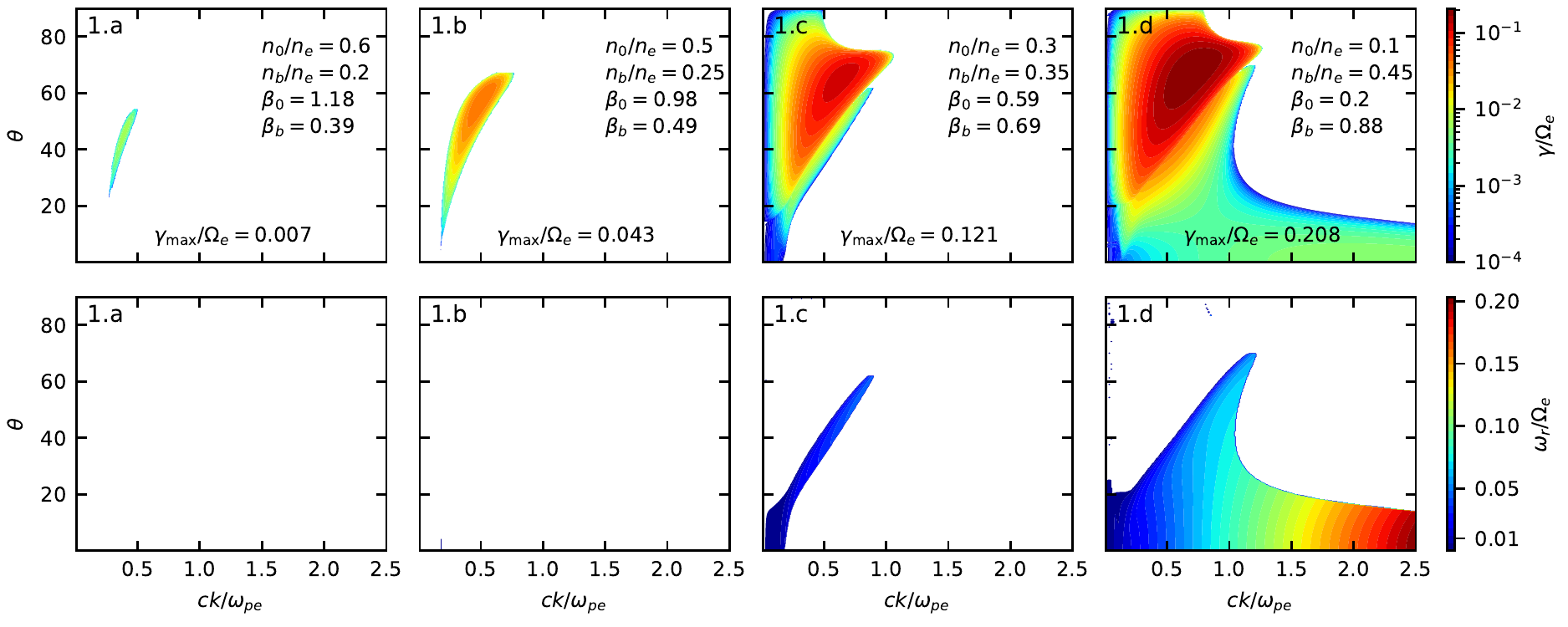}
    \caption{\label{f1} Color coded growth rates $\gamma/\Omega_e$ (top panels) and wave frequencies $\omega/\Omega_e$ (bottom panels) for cases 1.a, 1.b, 1.c and 1.d (from left to right).}
  \end{center}
\end{figure*}

\begin{figure*}[h!]
  \begin{center}
   \includegraphics[width=0.9\textwidth]{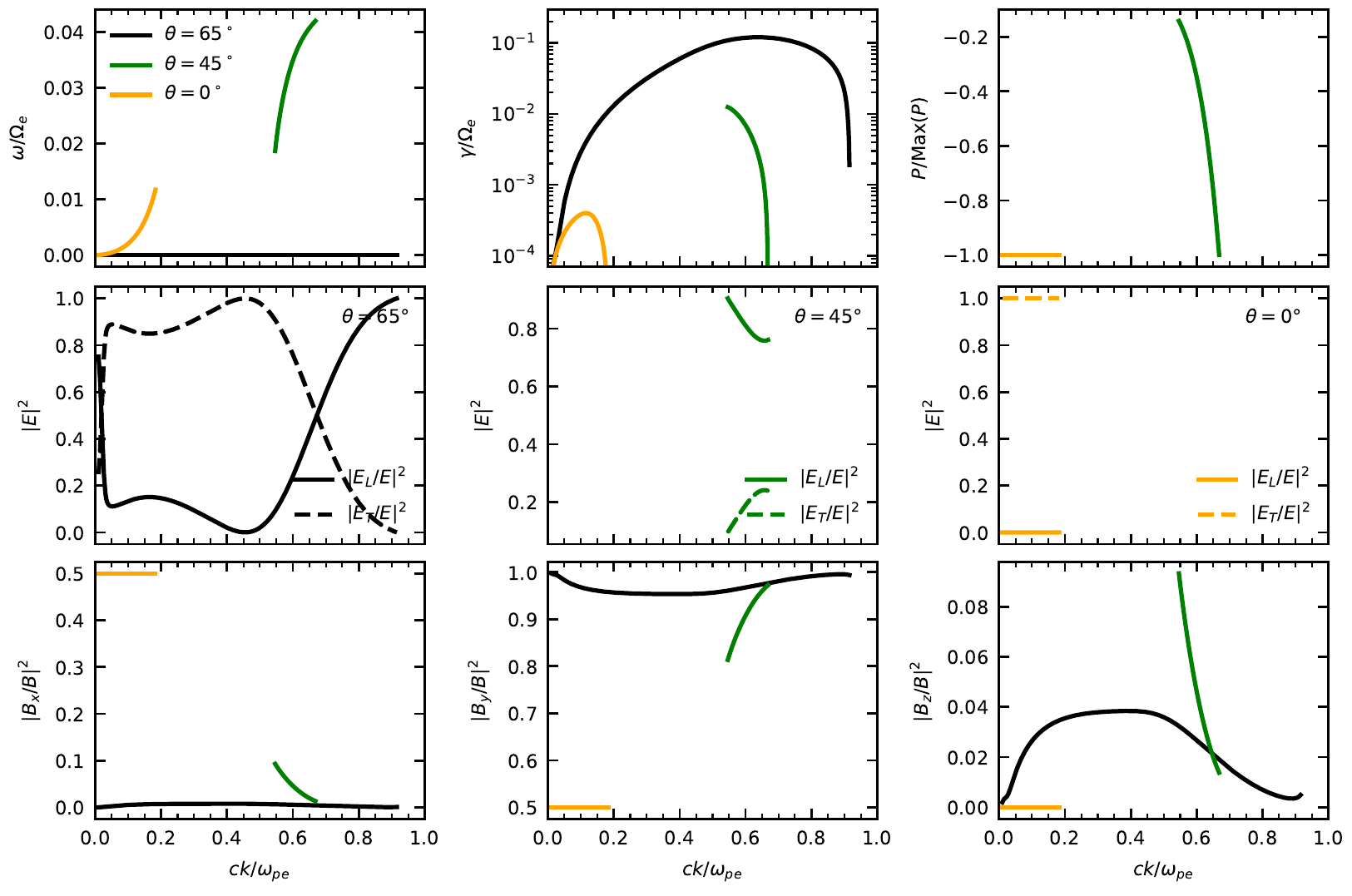}
\caption{\label{f2} The wave-number dispersion for the fastest growing unstable modes (with maximum growth rates) in case 1.c, corresponding to different angles of propagation ($\theta$) and different ranges of unstable wave-numbers: BEFI at $\theta = 65^{\rm o}$ (black), oblique HFI at $\theta = 45^{\rm o}$ (green) and FHFI at $\theta = 0^{\rm o}$ (orange). Top panels: wave frequency (left), growth rates (center), and polarization (right). Middle panels: longitudinal and transverse components of the wave electric components.  Bottom panels: cartesian components of the wave magnetic field. See also details in the text.}
  \end{center}
\end{figure*}

\begin{figure*}[t!]
  \begin{center}
    \includegraphics[width=0.9\textwidth]{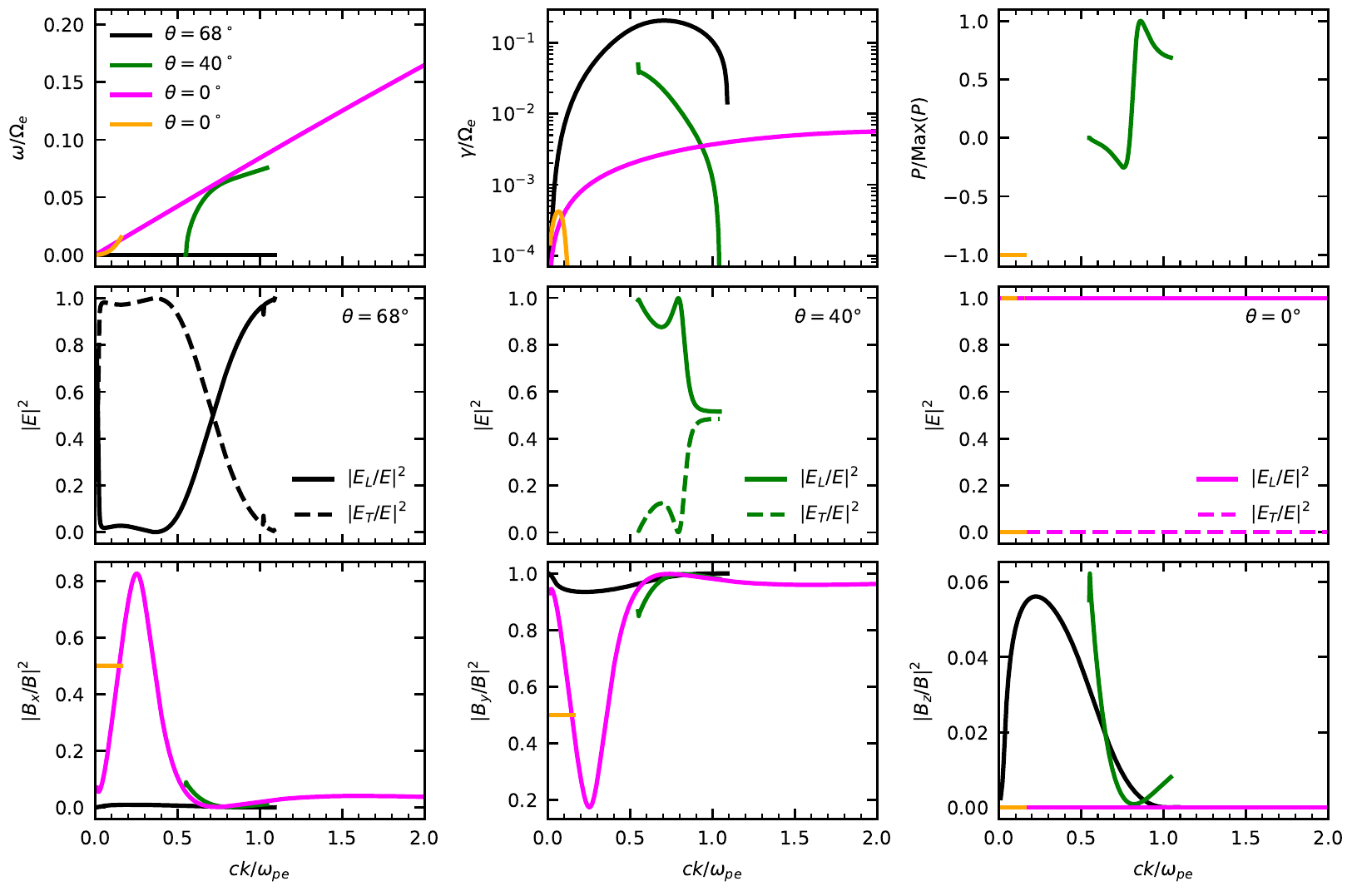}
\caption{\label{f3} The wave-number dispersion for the fastest growing unstable modes (with maximum growth rates) in case 1.d, corresponding to different angles of propagation ($\theta$) and different ranges of unstable wave-numbers: BEFI at $\theta = 68^{\rm o}$ (black), oblique HFI at $\theta = 40^{\rm o}$ (green), beaming ESI at $\theta = 0^{\rm o}$ (purple), and FHFI at $\theta = 0^{\rm o}$ (orange). Top panels: wave frequency (left), growth rates (center), and polarization (right). Middle panels: longitudinal and transverse components of the wave electric components.  Bottom panels: cartesian components of the wave magnetic field. See also details in the text.}
  \end{center}
\end{figure*}

\begin{figure*}[t!]
  \begin{center}
    \includegraphics[width=0.9\textwidth]{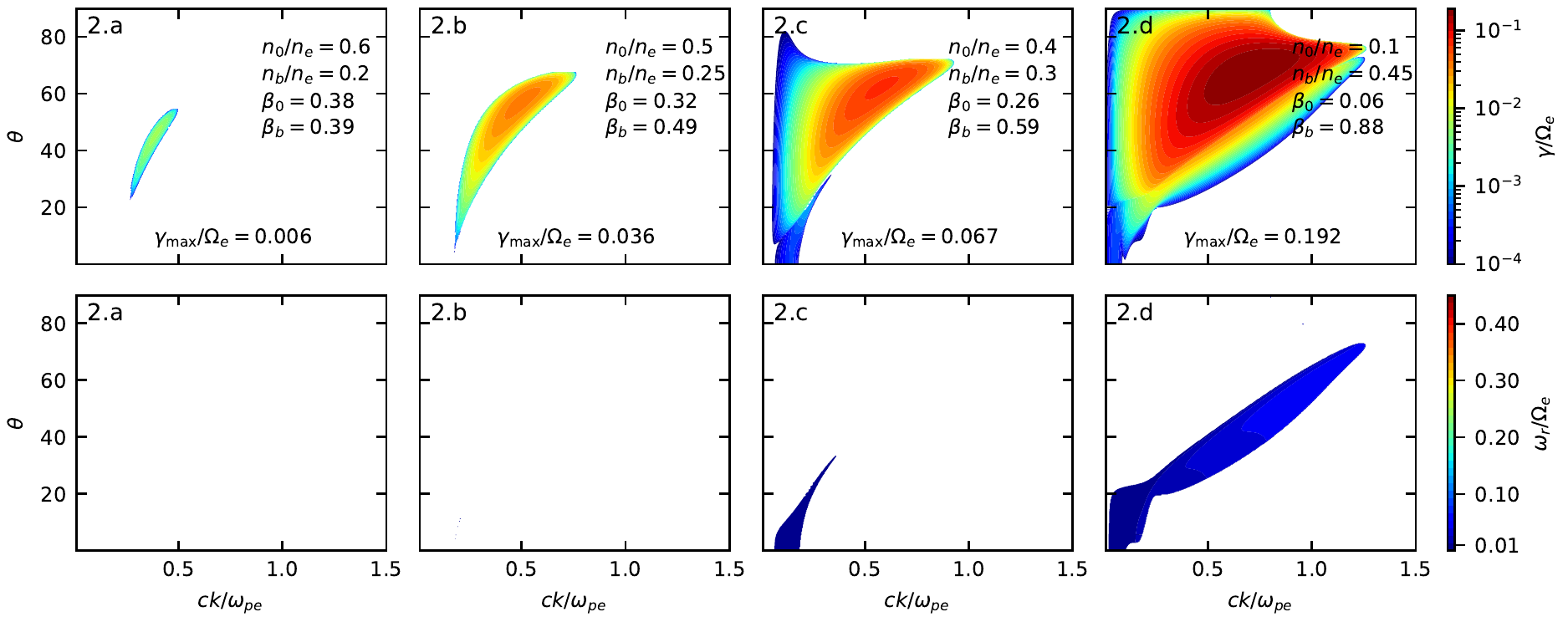}
    \caption{\label{f4}  Color coded growth rates $\gamma/\Omega_e$ (top panels) and wave frequencies $\omega/\Omega_e$ (bottom panels) for cases 2.a, 2.b, 2.c and 2.d, from left to right.}
  \end{center}
\end{figure*}

\begin{figure*}[t!]
  \begin{center}
    \includegraphics[width=0.9\textwidth]{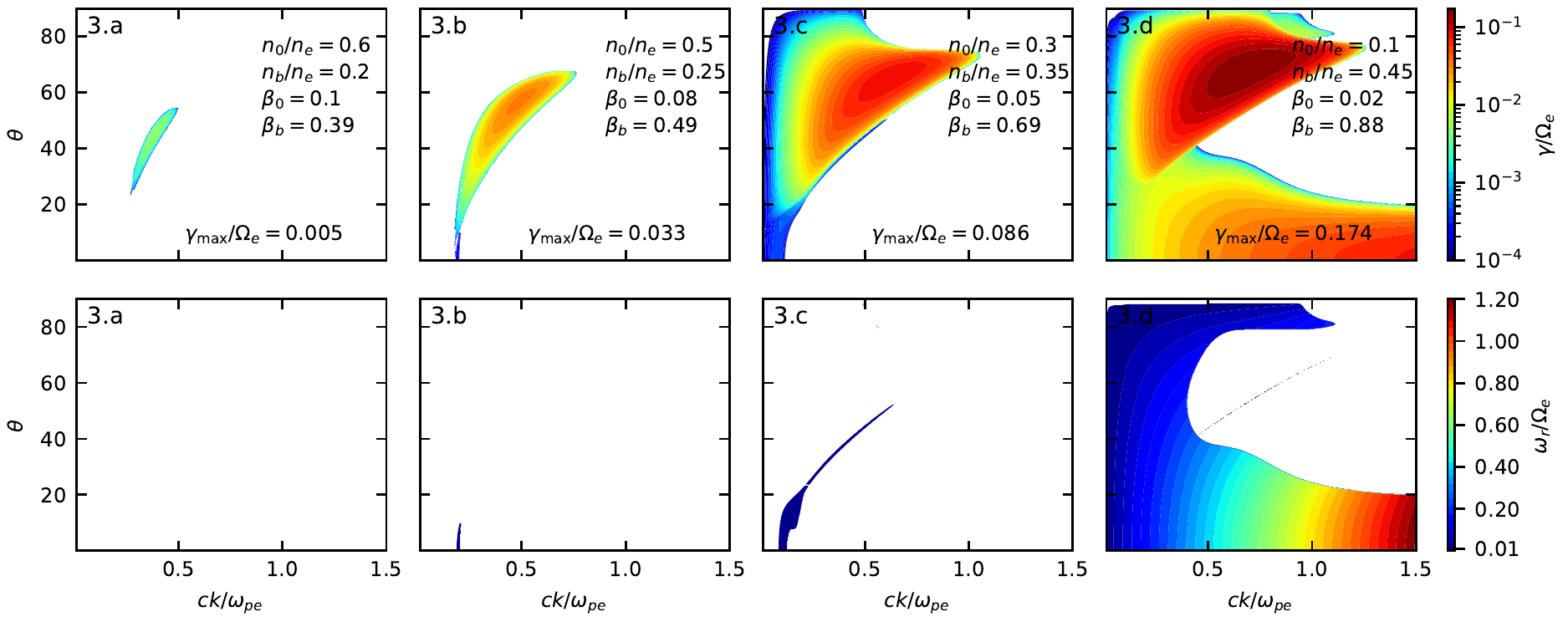}
    \caption{\label{f5} Color coded growth rates $\gamma/\Omega_e$ (top panels) and wave frequencies $\omega/\Omega_e$ (bottom panels) for cases 3.a, 3.b, 3.c and 3.d, from left to right.}
  \end{center}
\end{figure*}

\begin{figure*}[t!]
  \begin{center}
    \includegraphics[width=0.9\textwidth]{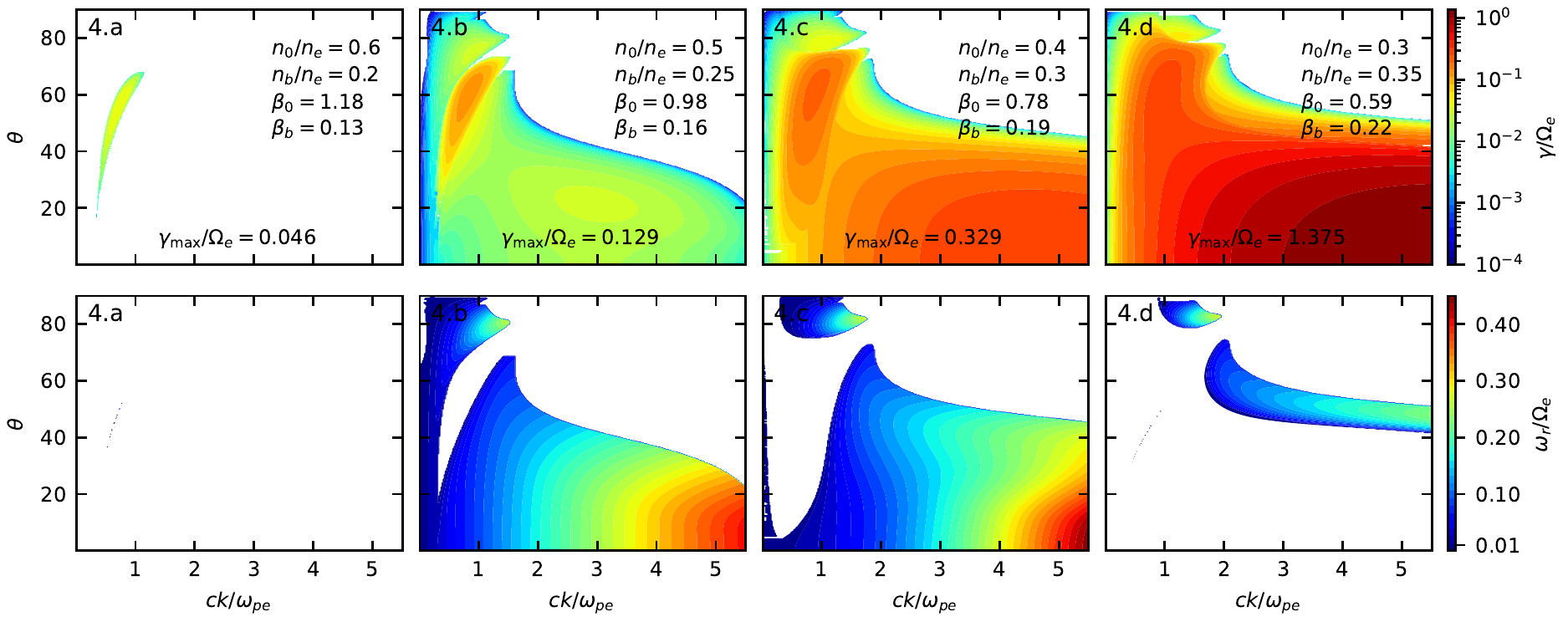}
    \caption{\label{f6} Color coded growth rates $\gamma/\Omega_e$ (top panels) and wave frequencies $\omega/\Omega_e$ (bottom panels) for cases 4.a, 4.b, 4.c and 4.d, from left to right.}
  \end{center}
\end{figure*}

\begin{figure*}[t!]
  \begin{center}
    \includegraphics[width=0.9\textwidth]{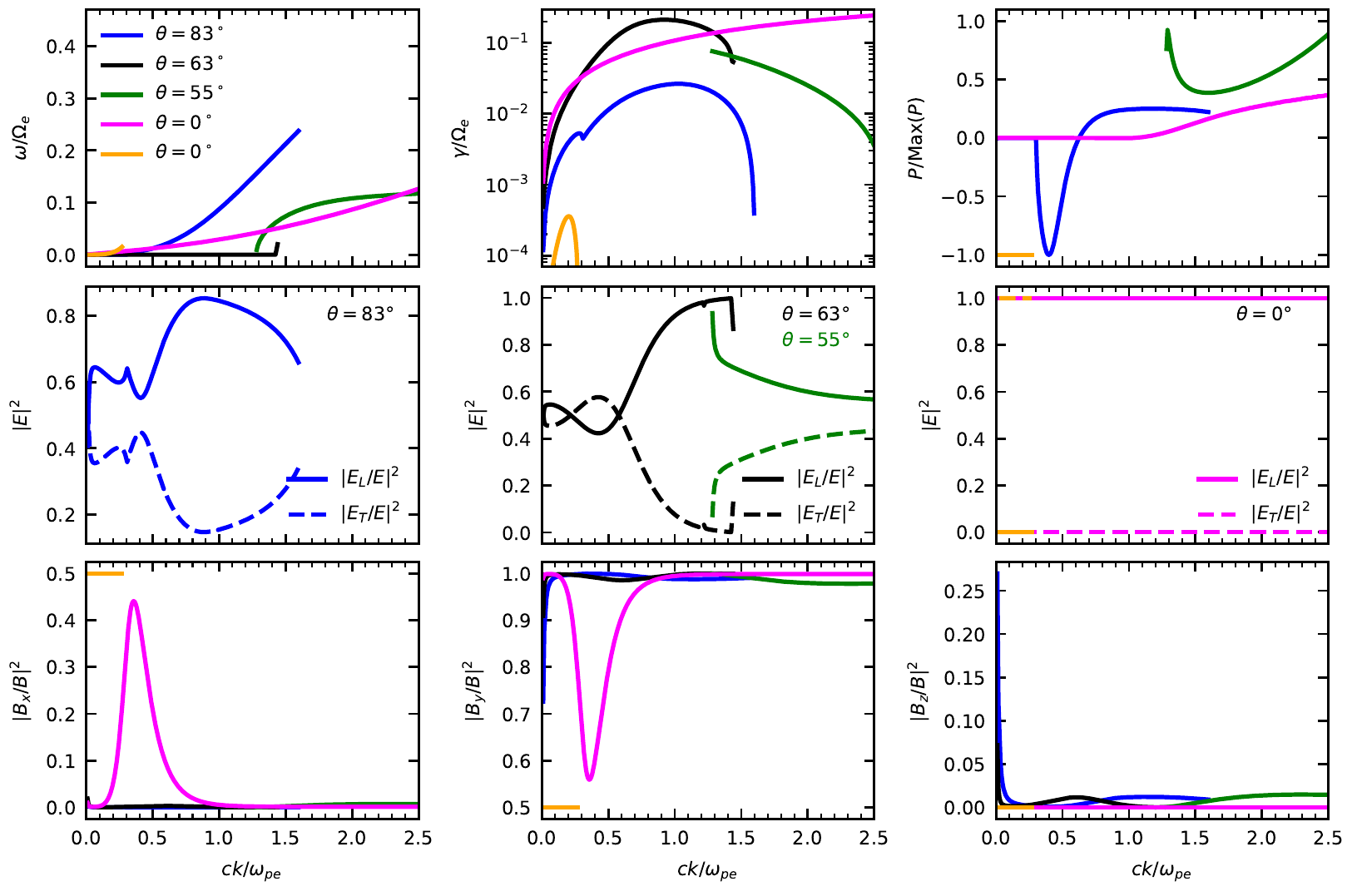}
    \caption{\label{f7} The wave-number dispersion for the fastest growing unstable modes (with maximum growth rates) in case 4.c, corresponding to different angles of propagation ($\theta$) and different ranges of unstable wave-numbers: BEFI at $\theta = 63^{\rm o}$ (black), oblique HFI-1 at $\theta = 55^{\rm o}$ (green), oblique HFI-2 at $\theta = 83^{\rm o}$ (blue), beaming ESI at $\theta = 0^{\rm o}$ (purple), and FHFI at $\theta = 0^{\rm o}$ (orange). Top panels: wave frequency (left), growth rates (center), and polarization (right). Middle panels: longitudinal and transverse components of the wave electric components.  Bottom panels: cartesian components of the wave magnetic field. See also details in the text.}
  \end{center}
\end{figure*}

Displayed in Figs.~\ref{f1}--\ref{f7} are the results for all these configurations, that allow us to delimit the specific regimes of this instability: (i) Near the instability thresholds, e.g., cases 1.a, 2.a, 3.a and 4.a, the (maximum) growth rates are very small, approaching marginal stability, i.e., $\gamma \to 0$, but, luckily, there is no other instability predicted by the theory in competition with BEFI. 
(ii) Regimes when additional instabilities can be identified in the wave spectra, but against which BEFI remains the dominant unstable mode, i.e., with the highest (maximum) growth rates. Overall, these results should show us how this instability is influenced by the background plasma, both by the relative density, $n_0/n_e$, and the thermal velocity, $\alpha_0/c$, of the background electrons. In Figs.~\ref{f1}, \ref{f4}, \ref{f5} and \ref{f6}, the white background signifies levels below the minimum level in the color bars (on the right) used to quantify the growth rate or wave frequency.


Fig.~\ref{f1} presents the four sub-cases of case 1, when thermal velocities of the beaming and background populations are comparable $\alpha_b/c \simeq \alpha_0/c = 0.07$, and are only slightly higher than beaming speed $U_b/c = 0.065$. 
The unstable spectra show the wave number ($ck/\omega_{pe}$) dispersion of the growth-rate ($\gamma/\Omega_e$, upper panels) and the wave frequency ($\omega_r/\Omega_e$, lower panels), as a function of the propagation angle ($\theta$). 
Both the growth rate and wave frequency are color coded on the right side of the respective panels. 
Very low growth rates of BEFI are obtained in case 1.a (left panels), with, e.g., a maximum $\gamma_{\rm max}/\Omega_e = 0.007$, when background electrons have a major density $n_0/n_e = 0.6 > n_b/n_e = 0.2$. 
With decreasing the density contrast between background and beaming electrons the peaking growth rate of BEFI increases with one order of magnitude for case 1.b, i.e., for $n_0/n_e =0.5$ and $n_b/n_e=0.25$, and may reach $\gamma_{\rm max} / \Omega_e \simeq 0.21$ for case 1.d, for even more dense electron beams with $n_b/n_e = 0.45 > n_0/n_e = 0.1$. 
The results from case 1.d, are indeed very similar to those obtained by \cite{Lopez-etal-2020b} in the absence of background electrons. 
The comparison of the four cases shows a clear and significant inhibition of BEFI under the influence of background electrons.
This instability remains aperiodic ($\omega_r = 0$) for all these plasma configurations, as a specific feature that may help to differentiate from other unstable modes. 

In the unstable spectra for cases 1.c and 1.d, we can also distinguish other modes of different nature, in general with finite wave frequency $\omega_r \ne 0$, but (maximum) growth rates much lower than those of BEFI. Distinction can still be made between these two spectra. Thus, in panels 1.c, at small quasi-parallel angles $\theta < 20^{\rm o}$ and small wave-numbers, we may identify the firehose heat-flux instability (FHFI), and for more oblique angles and larger wave-numbers the oblique branches of heat-flux instabilities, which may combine FHFI and whistler heat-flux (WHF) instabilities, discussed to more detail in \cite{Lopez2019a}, see, e.g., their Fig.~3. 
These heat-flux instabilities are triggered by an effective anisotropy in velocity space, as resulted from the asymmetry of thermal spread of the beaming and background populations. These instabilities may therefore not be very sensitive to the variation of (relative) number density.
For more dense counter-beams in case 1.d, the electrostatic beaming instabilities (i.e., with a major longitudinal electric field component $E_L = {\bf E} \cdot {\bf k}$) are also predicted at parallel and small $\theta$ propagation.
In this case the quasi-parallel unstable spectrum becomes already complex, showing a superposition of unstable modes with finite frequency, increasing with the wave number. 
However, the maximum growth rates of all the other unstable modes remain much lower than those of BEFI. Clearly, in all these cases BEFI is not competed by other instability, and it is solely predicted to operate as the main radiative mechanism, with possible consequences on the relaxation of electron counter-beams. The corresponding plasma beta parameters, $\beta_0$ and $\beta_b$, see Table~\ref{t1}, take comparable values, around or slightly lower than 1, which means conditions near the equipartition of kinetic and magnetic energy, specific to the solar wind, in, e.g., CIRs and terrestrial bow-shock. 

Aiming to decipher the nature of the unstable modes, in Fig.~\ref{f2} we describe the main properties of the fastest growing modes, i.e., those with maximum growth rates, corresponding to each instability in case 1.c from Fig.~\ref{f1}. 
Upper panels in Fig.~\ref{f2} show the wave frequency (left), the growth rate (middle), and the polarization (right) defined as $P = \text{Sign}(\omega_r) \text{Re} \left\{i(E_x/E_y)\right\}$. 
This polarization is relevant for the electromagnetic modes, circular (or eliptically) polarized, $P> 0$ meaning right-handed (RH) polarization and $P < 0$ left-handed (LH) polarization. %
The maximum growth rate ($\gamma_{\rm max}/\Omega_e = 0.208$), associated with the fastest growing mode, is obtained for BEFI at $\theta \simeq 65^\circ$ (black lines), as an aperiodic mode ($\omega_r=0$) purely growing in time. 
Lower panels in the middle row show components of the wave electric field, longitudinal (or parallel) and transverse to the direction of propagation (i.e., to ${\bf k}$), for three distinct modes: BEFI with maximum growth rate at $\theta = 65^{\rm o}$ (black lines in the left panel); the oblique branch of the HFIs, in this case, firehose-like modes, LH-polarized ($P < 0$), and with maximum growth rate at $\theta = 45^{\rm o}$ (green lines in the middle panel); and for parallel propagation ($\theta = 0$) a FHFI, circularly LH-polarized with $P = -1 < 0$ and only a transverse component ($E_T$) of the wave electric field (orange lines in the right panel).
Shown in the bottom panels are the corresponding cartesian components of the wave magnetic fields, which confirm the nature of these modes. Notice that for BEFI the major component is $B_y$, which is another common feature with the firehose instability driven by the temperature anisotropy of electrons \citep{Camporeale2008}. 

Fig.~\ref{f3} shows the same details as in Fig.~\ref{f2}, but for the main properties of the fastest growing unstable modes in case 1.d, those  corresponding to the peaking growth rates in Fig.~\ref{f1}. 
In this case the relative density of the background electrons is only $n_0/n_e =0.1$, and the unstable wave spectra resemble those obtained by \cite{Lopez-etal-2020b}, for similar plasma parameters but in the absence of background electrons. 
For BEFI (black lines) the value of maximum grow rate is higher, and is obtained at $\theta \simeq 68^\circ$. The maximum growth rate of the oblique HFI (green lines) remains lower than that of BEFI, and is obtained at $\theta \simeq 40^\circ$. 
But in this case, in the oblique HFI one may observd that FHF (LH-polarized, with $P <0$, at lower wave-numbers) couples with WHF (RH-polarized, with $P > 0$, at higher wave-numbers), as also shown by \cite{Lopez-etal-2020a}. 
It should also be remarked the similarity of the properties of these modes with those obtained for an asymmetric electron plasma-beam system \citep{Lopez-etal-2020a}. 
For parallel propagation ($\theta = 0^\circ$) we find not only the FHFI (orange lines), but also the electrostatic (ES) electron beaming instability (EBI, with purple lines). For both of them maximum growth rates are less than that of BEFI. 
This oscillatory ($\omega \ne 0$) beaming mode is most probably excited by the asymmetric counter-drifting beam and background populations of electrons, by contrast to previous studies in the absence of background electrons \citep{Lopez-etal-2020b}, where symmetric counter-beams were able to trigger an aperiodic ($\omega = 0$) two-stream instability. 
This ES mode seems to couple to EM modes, FHF modes with a $B_x$ component at low wave-numbers, and the other oblique, BEFI or WHF modes, with a major $B_y$ transverse component of the wave magnetic field. Apparently with a hybrid nature, this mode is not of interest in our present analysis, but could be investigated in future studies.

Fig.~\ref{f4} presents the unstable solutions obtained for cases 2.a - 2.d, similar to cases 1.a - 1.d, but for a cooler background population, this time with $\alpha_0/c = 0.04 < \alpha_b/c = 0.07$. 
For the same relative densities, see Table~\ref{t1}, profiles of the unstable spectra are similar to those obtained in Fig.~\ref{f1}, showing a dominance of BEFI. The highest peaking (maximum) growth rates are obtained for BEFI, in general, at oblique angles, which increase with lowering the influence of the background electrons (from left to right). The growth rates increase the same way, but their maximum values, indicated in each panel in Fig.~\ref{f4}, are lower than those obtained in Fig.~\ref{f1}, meaning that BEFI is inhibited by a cooler background population of electrons. Note also, that unlike case 1.d., the ES instabilities are missing from the unstable spectra of case 2.d, despite the similarity between the plasma configurations.

\begin{figure*}[h!t!]
  \begin{center}
    \includegraphics[width=0.9\textwidth]{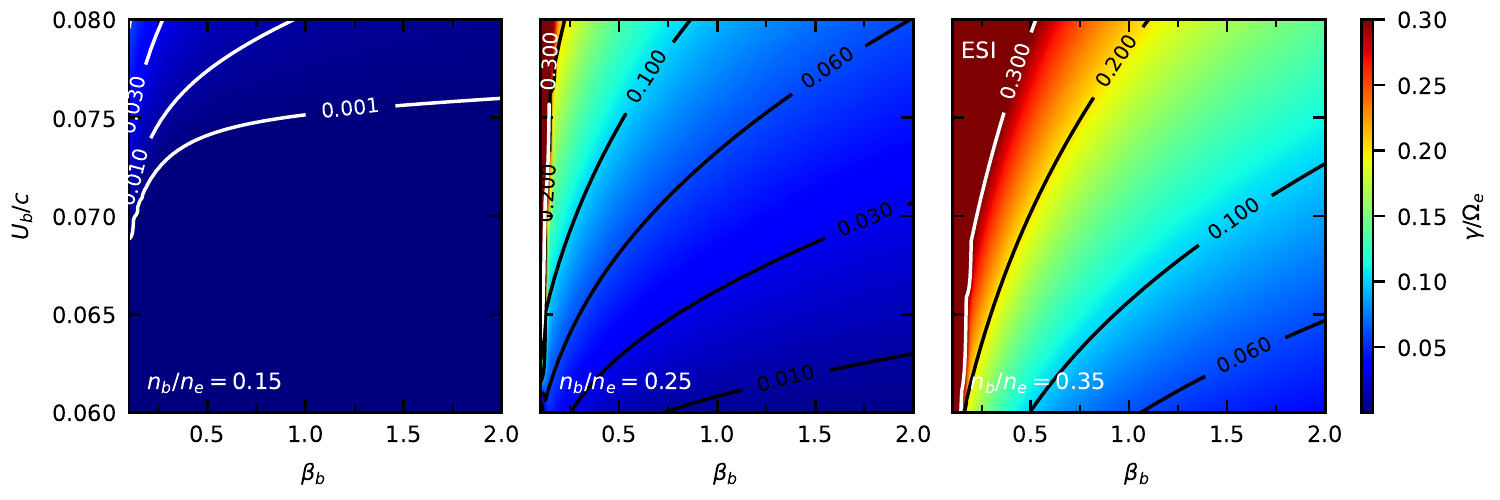}
    \caption{\label{f8} Maximum growth rates $\gamma_\text{max}/\Omega_e$ (color coded) obtained for BEFI for $\alpha_0=\alpha_b$ (cases 1), and different density ratios $n_b/n_e=0.15$ (left), $n_b/n_e=0.25$ (center), and $n_b/n_e=0.35$ (right). Levels above contours at $\gamma_{\rm max}/\Omega_e=0.3$ correspond to the ESI (middle and right panels).}
  \end{center}
\end{figure*}


\subsection{From BEFI to ES instabilities: cases 3 and 4}

Next let us see the properties of BEFI for cases 3.a - 3.d, when the electron background population is even cooler, i.e., $\alpha_0/c = 0.02$, and for the same set of relative number densities, see Table~\ref{t1}. 
The unstable solutions are displayed in Fig.~\ref{f5}, i.e., color coded wave frequencies (top panels) and growth rates (bottom panels). In this case, BEFI remains fairly distinct and dominant, with growth rates higher (or even much higher) than all the other modes predicted by the linear theory. 
The inhibiting effect of a cooler electron background is confirmed, by, e.g., the maximum growth rates, which are indicated in each panel, and are lower than those obtained for the corresponding cases in Fig.~\ref{f4}. 
However, in case 3.d, when the relative density of the background population is very low, i.e., $n_0/n_e = 0.1$, the growth rates of ES modes with (quasi-)parallel wave-vectors become important, they are still less than those of BEFI, but are already markedly higher than those of ES modes obtained in case 1.d. 
This is what we can call, as also suggested by \cite{Lopez-etal-2020b} and \cite{Moya-etal-2022}, the transition from the dominance of BEFI to the regime of ES instabilities, specific to much cooler electron populations, see, e.g., Fig.4 in \cite{Moya-etal-2022}.
Although these electrostatic instabilities are not the object of our study, we can explain these results by mentioning that in the velocity distributions (not shown here) case 3.d shows peaks of the counterbeams and corresponding slopes ($\delta f_b/\delta v \propto \gamma > 0$) more prominent than those for case 1.d. In case 2.d (and also case 1.c) the same peaks and corresponding slopes are much lower, below the threshold of these instabilities.

In Figs.~\ref{f6} and \ref{f7} we show that this transition can be even steeper when the electron beams are cooler than background population, i.e., for cases 4.a - 4.d., when $\alpha_b = 0.04 < \alpha_0 = 0.07$.
In Fig.~\ref{f6} growth rates of BEFI (top panels) show the same inhibition under the influence of background electrons, but contrary to that, BEFI keeps the highest growth rate only for sufficiently dense background population, for instance, in cases 4.a and 4.b. 
Already in case 4.b, but especially in the other two cases, 4.c and 4.d, the spectra of instabilities become much more complicated due to new unstable solutions, both at oblique propagation angles and in directions parallel and quasi-parallel to the magnetic field. 
By comparison with BEFI, these new instabilities are oscillatory (or periodic) in time, i.e. with $\omega_r \ne 0$, see the bottom panels in Fig.~\ref{f6}. 
This property helps us to differentiate them, given that their growth rates become comparable (case 4.c) or even exceed (case 4.d) those of BEFI. 
We should not forget that the two-stream aperiodic instability predicted in the absence of background population, is expected in this case as well. That seems to be identified in case 4.d at low angles and large wave-numbers, as the mode with highest (maximum) growing rates but with a very small frequency $\omega_r \to 0$.
For a better distinction, but also for a preliminary identification of the nature of the unstable modes, in Fig.~\ref{f7} we represent in detail the properties of the most unstable modes, associated with the maximum growth rates of different modes distinguished in case 4.c (as above in Figs.~\ref{f2} and \ref{f3}). 

The maximum growth rate of BEFI (black lines) is obtained at $\theta = 63^{\rm o}$, and in this case is comparable with that of ES electron beaming instability (at $\theta = 0^{\rm o}$), indicated with purple lines.
At oblique angles, this time we can also identify two heat-flow (HF) instabilities with $\omega_r \ne 0$. 
We know already the oblique HF found at intermediary oblique angles and larger wave-numbers, which is a whistler HF (WHF), with RH polarization  and a maximum growth rate at $\theta = 55^{\rm o}$), indicated with green lines as in Fig.~\ref{f3} in Fig. 7. 
A new unstable mode is predicted at very oblique angles (indicated with blue lines in Fig. 7), and combines a firehose HF (FHF) LH polarization, at low wave-numbers, with a WHF RH polarization, at large wave-numbers. 
Even the profile of growth rates shows two humps, corresponding to two different modes. 
The one obtained at large wave-numbers connects apparently to the WHF mode obtained at lower oblique angles, e.g., for all cases 4.b, 4.c and 4.d, but remains however distinct. For both cases 4.c and 4.d growth rates of these two modes remain lower than BEFI. 
The influence of these RH-polarized modes extends to low angles of propagation, becoming visible at large enough wave-numbers and explaining the major magnetic field component $B_y$ obtained already in case 1.d for the ES mode (purple lines). Note, however, that in case 4.d, by far the highest growth rate is that of the aperiodic two-stream instability propagating in parallel direction, a robust and highly competitive instability of two symmetric highly dense counter-beams, as described by \cite{Lopez-etal-2020b}.
For the oblique instabilities, the wave-number dispersion of the electric and magnetic field components shows similar profiles, and all resemble those obtained in Fig.~\ref{f3}. 
By comparison to case 1.d, BEFI has already a significant $E_L$ at low wave-numbers, where $E_L \sim E_T$. 
However, the fastest growing mode, with maximum growth rate, has the same hybrid nature with a major $E_L >> E_T$, and a major $B_y$. 
For $\theta = 0^{\rm o}$ we again find a purely electromagnetic FHF (orange lines) with LH polarization and a growth rate much lower than that of BEFI.


\subsection{Maximum growth rates (thresholds) of BEFI}

Linear theory can also offer a more comprehensive image of BEFI, if we compute the maximum growth rates and build maps of their contour levels as a function of the main plasma parameters, in this case, the (normalized) beam speed $U_b/c$, i.e., the main source of free energy, but also the plasma beta parameter, e.g., $\beta_b$ for the beam.
Fig.~\ref{f8} displays the normalized maximum growth rates $\gamma_{\rm max}/\Omega_e$, coded according to the color bar on the right side and obtained for BEFI for the situations specific to case 1, when the beam and background electrons have the same thermal spread $\alpha_b = \alpha_0$, and three different density ratios $n_b/n_e=0.15$ (left), $n_b/n_e=0.25$ (center), and $n_b/n_e=0.35$ (right). 
The main features of BEFI are already known, i.e., for the same $\beta_b$ the growth rates are significantly enhanced with increasing the beam/drift speed. 
Contour lines (black or white) can be fitted to various mathematical expressions of $U_b/c$  as a function of $\beta_b$, see e.g., in \cite{Shaaban-etal-2018-pop} and \cite{Moya-etal-2022}, to quantify the beam speed thresholds of this instability, though here we limit to a qualitative analysis. 

The effect of background electrons becomes also obvious if we compare with the maximum growth rates of BEFI obtained in \cite{Lopez-etal-2020b} for $n_0/n_e = 0$, which are markedly higher than those derived here, e.g., in the right panel for the same beam/drift velocity and same plasma beta.
Moreover, the three panels in Fig.~\ref{f8} show a uniform effect of the background population, which tends to suppress the instability, markedly inhibiting (from right to left) the maximum growth rates, and increasing the instability thresholds, see, e.g., contours at $\gamma_{\rm max}/\Omega_e =$ 0.06 and 0.1 in the middle and right panels. 
In the middle panel, for a relative density of the beam $n_b/n_e = 0.25$, and $n_0/n_e = 0.5$ for the background, one can observe that BEFI can still be triggered with a reasonable maximum growth rate $\gamma_{\rm max}/\Omega_e \simeq [0.10-0.15]$, if the beta parameter and beam speed are sufficiently high \footnote{Higher values of $U_b$ must be considered with caution to not exceed the non-relativistic limit of our approach $U_b/c \lesssim 0.4$ (where $c$ is the speed of light in vacuum), above which the electrons with energy $E > 100$ keV are weakly relativistic.}, respectively, $\beta_b > 0.2$ and $U_b/c > 0.065$. 
If the background electrons are dominant, e.g., with a relative density $n_0/n_e = 0.7$, e.g., in the left panel, BEFI can be barely excited, with very low growth rates $\gamma_{\rm max}/\Omega_e \simeq 10^{-3}$ approaching and describing the plasma conditions of marginal stability ($\gamma \to 0$) against BEFI. 
With decreasing the presence of background electrons, the beam speed characteristic to marginal stability is also markedly lowered, as already found for the instability thresholds. 
On the other hand, in the middle and right panels, above the contour level around $\gamma_{\rm max}/\Omega_e \simeq 0.3$ we can identify the regime of ESI, whose maximum growth rates become much superior to BEFI. 

The shape of these thresholds is very similar to the one obtained for the thresholds of firehose heat-flux instability (FHFI) induced in the direction parallel to the magnetic field by a single (asymmetric) strahl/beam in the solar wind \citep{Shaaban2018, Shaaban-etal-2018-pop}. 
By virtue of these properties, we can treat BEFI as an instability triggered by a double heat-flux. But more than that, BEFI is from the category of the oblique heat-flux instabilities, that propagate/develop at highly oblique direction with respect to the magnetic field, as the oblique whistler heat-flux instability \citep{Verscharen-etal-2019,Lopez-etal-2020a}. 
By contrast with the parallel heat-flux instabilities, the oblique ones can effectively contribute to the relaxation of the electron beams, through an efficient resonant scattering of beaming electrons (and do not require that the electrons and waves counter-propagate), as shown not only in numerical simulations \citep{Micera-etal-2020,Vo-Cattell-2022} but also in a series of recent observations \citep{Cattell-etal-2020}. 
Therefore, we expect that BEFI-like instabilities play an effective role in the relaxation of double electron strahls/beams, those counterbeams with a sufficiently high thermal spread, as predicted by their linear proprieties discussed in this section. 
This could be the case of electron counterbeams observed in CIRs, but also in the interplanetary shocks and CME foreshocks at sufficiently large heliocentric distances (e.g., 1~AU and beyond).  
We do mention, however, that the observed electron counterbeams are not necessarily symmetrical, in which case the oblique instability can change its properties, becoming periodic ($\omega_r \ne 0$) and possibly whistler-like in nature.


\begin{figure}[h!]
  \begin{center}
    \includegraphics[width=0.45\textwidth]{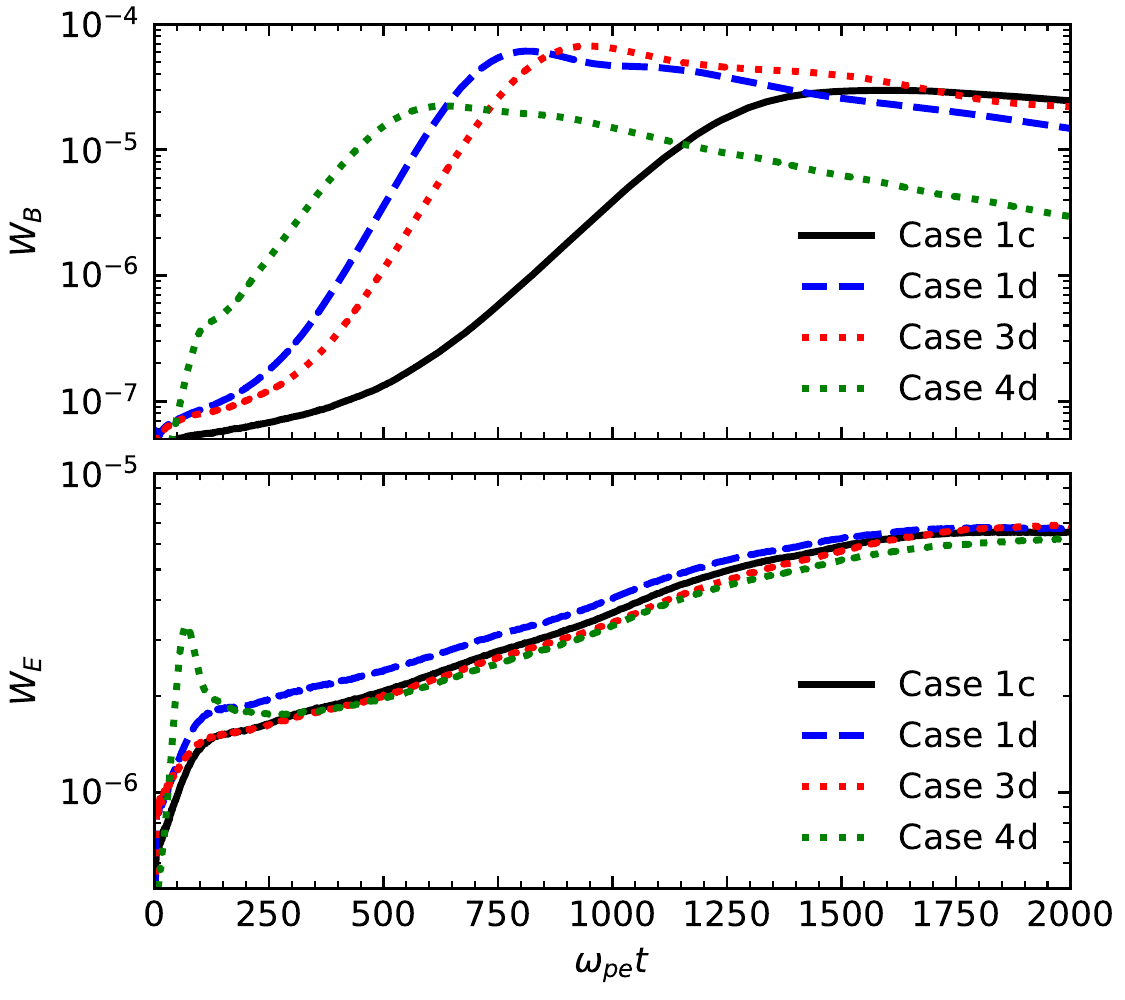}
    \caption{\label{f9} Temporal evolution of the magnetic (top) and electric (bottom) field energy, for cases 1.c, 1.d, 3.d and 4.d.}
  \end{center}
\end{figure} 

\begin{figure*}[h!]
  \begin{center}
    \includegraphics[width=0.7\textwidth]{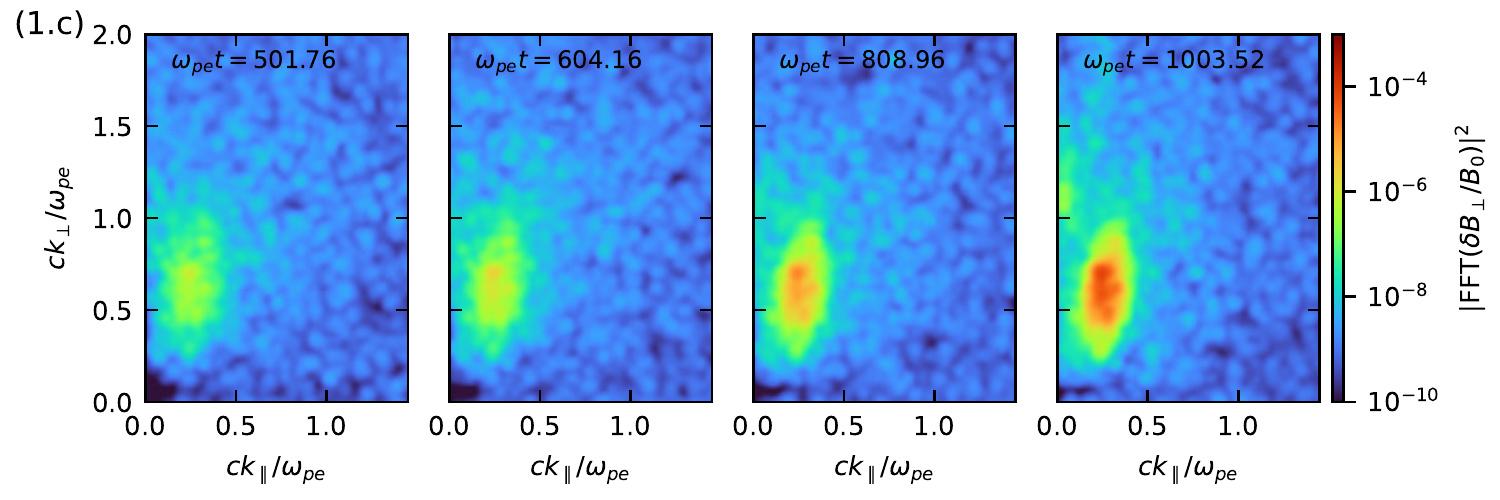}    
    \includegraphics[width=0.7\textwidth]{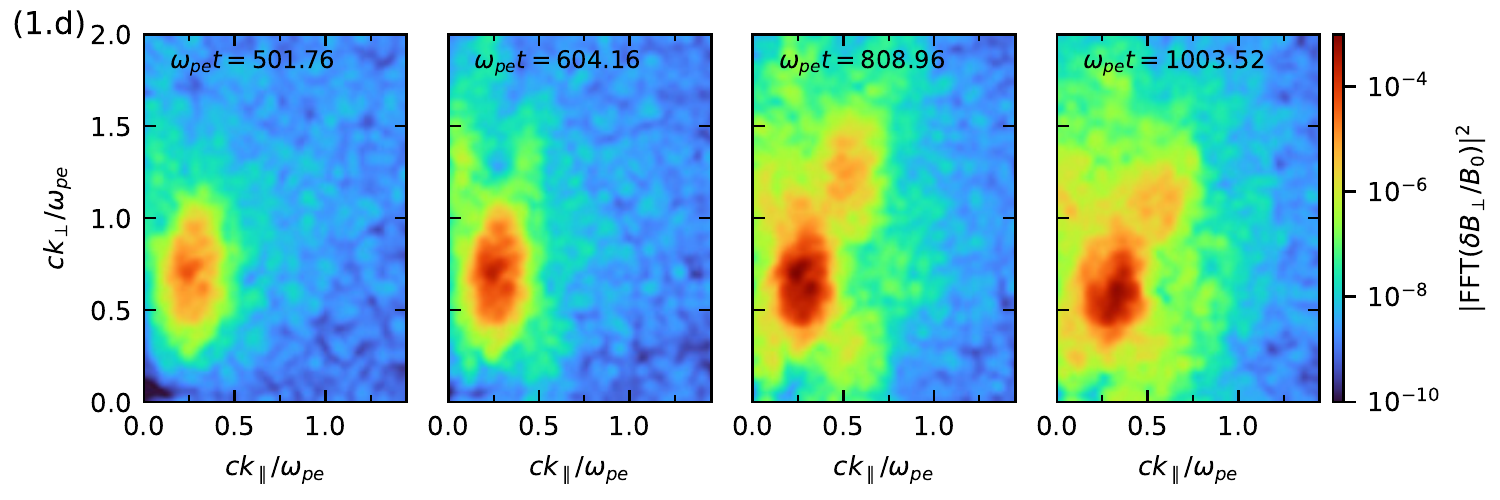}
    \includegraphics[width=0.7\textwidth]{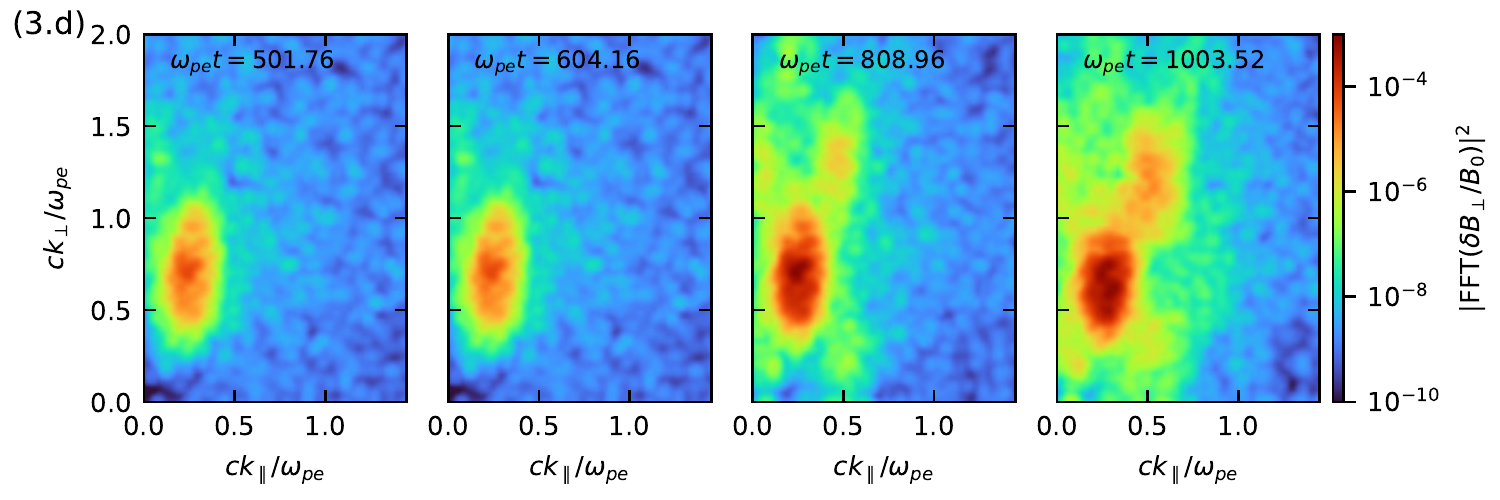}
    \includegraphics[width=0.7\textwidth]{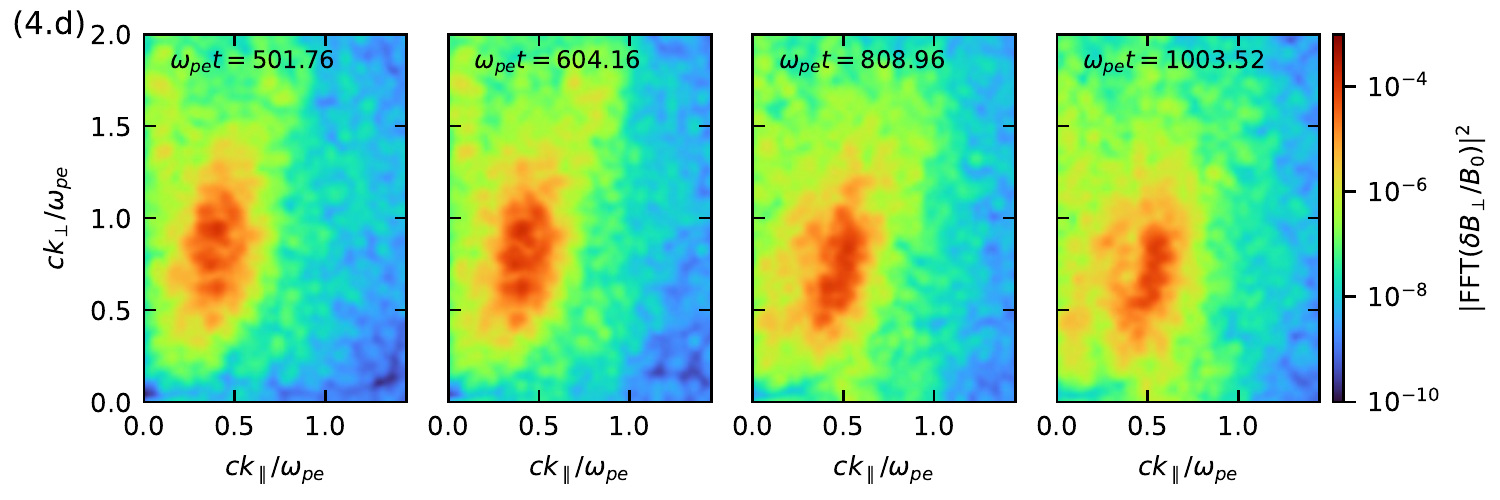}
    \caption{\label{f10} Time snapshots of the FFT (normalized) energy density computed for the out of plane (perpendicular) component of the magnetic field, for cases 1.c, 1.d, 3.d, and 4.d.}
  \end{center}
\end{figure*} 

\begin{figure*}[h!]
  \begin{center}
    \includegraphics[width=0.7\textwidth]{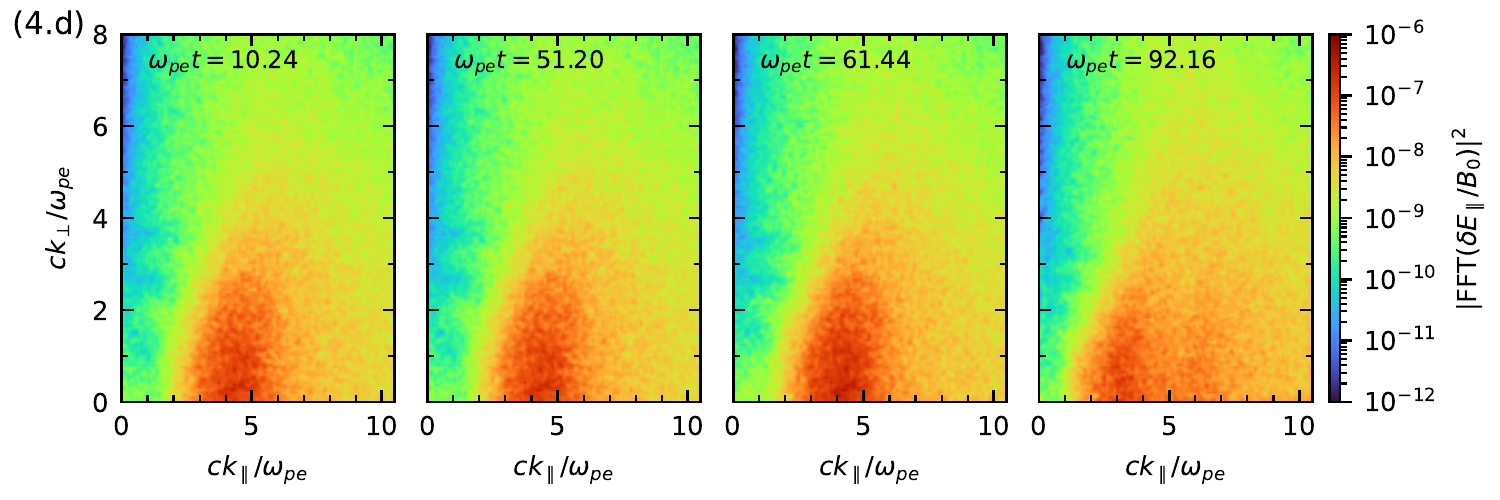}
    \caption{\label{f11} Time snapshots of the FFT (normalized) energy density computed for the parallel electric field component in case 4.d, showing an early time fast growth and saturation of ES instability.}
  \end{center}
\end{figure*}

\section{Particle-in-cell simulations}\label{pic}

In order to validate the predictions from linear theory examined in section \ref{befi}, here we present results from simulations which also describe the evolution of BEFI in time. 
We have used a 2D explicit PIC code based on the KEMPO1 code from \cite{Matsumoto1993}. 
Our simulation domain is composed by $512\times512$ grid cells, with $L_x=L_y=153.6\,c\omega_{pe}$ and 625 particles per grid per species. The mass ratio is $m_p/m_e=1836$, the plasma to gyro-frequency $\omega_{pe}/\Omega_{ce}=20$, the time step is $\Delta t=0.01/\omega_{pe}$ and the simulation runs until $t_\text{max}=2000/\omega_{pe}$. 
We chose to present the simulation results for four cases, 1.c, 1.d, 3.d and 4.d, which confirm the excitation of BEFI for different plasma conditions, but also allow us to compare the wave fluctuations triggered by different initial conditions. 
That is either for different relative densities of the electron beams, if we compare cases 1.c and 1.d, or for different thermal speeds of the electron populations, contrasting cases 1.d and 3.d, or 1.c and 4.d.

Fig.~\ref{f9} displays the evolution in time of the fluctuating magnetic field energy density $W_B = \int \delta B^2/B_0^2\, dx dy$ and the electric energy density $W_E = \int \delta E^2/B_0^2\, dx dy$, for the time interval of the simulations. 
From the figure we can see that cases 1.c, 1.d and 3.d (all cases initially satisfying $U_b \lesssim \alpha_b$) are qualitatively similar. 
In all three cases, in agreement with linear theory predictions, the fastest developing BEFI (i.e., with maximum growth rate) has a hybrid nature, with an electric field component (mainly contributing to $W_E$) that grows at the beginning faster than the electromagnetic (EM) transverse  component ($W_B$). 
However, as time advances the EM energy density $W_B$ arises and reaches levels of about one order of magnitude larger than the electric energy density $W_E$. 
For cases 3.d and 4.d the increasing slopes of $W_B$ are indeed higher than case 1.d, as predicted by the maximum growth rates ($\gamma_{\rm max}$) obtained from linear theory \footnote{A direct correspondence of these slopes with $2\gamma_{\rm max}$ cannot be done because not only the mode with maximum growth rate develops with the increase of time.}. 

The growth of BEFI fluctuations in time is confirmed in Fig.~\ref{f10} by the FFT spectra of the normalized energy density $|{\rm FFT}(\delta B/B_0)|^2$, computed for the out of plane (perpendicular) component of the fluctuating magnetic field. The levels of fluctuations are coded in the right-hand color bars. 
Displayed are four time snapshots up to (or near) the saturation, for the same cases 1.c, 1.d, 3.d and 4.d, from top to bottom, respectively. The 2D dispersion at large propagation angles in the wave-vector space ($k_\parallel, k_\perp$) resemble those from linear theory, especially at early moments in time, when BEFI fluctuations do not yet reach very high amplitudes (intensities) to be affected by the nonlinear decays. 
Additional spots that are visible later in time at different propagation angles, may indeed signify fluctuations of daughter waves generated nonlinearly via three- or four-waves nonlinear decays. 
These results are very similar to Fig.~5 in \cite{Lopez-etal-2020b}, obtained for BEFI in the case with no electron background. 
However, BEFI fluctuations are visibly inhibited by the presence of background electrons. 
In this sense, the contrast between the levels of fluctuations in Figure 10 is also very relevant, such as those obtained for the same time snapshots in cases 1.c and 1.d.  

In case 4.d (with $U_b > \alpha_b$), our BEFI is predicted by linear theory in close competition with the electrostatic (ES) instabilities. 
(See also the results presented in Figs. 9, 10 and 11 in \citet{Lopez-etal-2020b}, where the initial conditions also considered beam speeds higher than thermal speeds, but in the absence of an electron background.) 
Indeed, the green dotted-line in Fig.~\ref{f9} shows a quick increase and relaxation of the fluctuating electric energy density $W_E$, with a narrow and not very high peak, followed by a drop and then by a more robust growth of the magnetic energy density $W_B$ due to BEFI. 
In this case primary excited is the ES instability, at much lower time scales, as already indicated in Fig.~\ref{f9}. BEFI develops as a secondary but more robust instability, and it is also confirmed in Fig.~\ref{f10} last row, for the same time scales of BEFI in cases 1.c, 1.d and 3.d. 
However, for case 4.d, the oblique maxima of BEFI are more disperse or less compact, most probably, due to linear or nonlinear interactions with fluctuations of other nature. 
The earlier ES excitations propagating at small angles with respect to the magnetic field are confirmed in Fig.~\ref{f11}, where we display earlier time snapshots of the FFT (normalized) energy density for the parallel electric field component in case 4.d. 
The levels of fluctuations are color coded in the right-hand bars, and reach a maximum (saturation) at about $\omega_{pe} t = 61.44$, much earlier than the first time snapshot shown in Fig.~\ref{f10}. 


\section{Conclusions} \label{conc}
Since space plasmas are weakly collisional (or even non-collisional), we expect wave instabilities to have multiple implications, especially by facilitating the conversion of free energy of plasma particles, as well as energy transfer between species. 
\cite{Lopez-etal-2020b} and \cite{Moya-etal-2022} have recently shown that two symmetric electron counter-beams, aligned to the guiding magnetic field, can induce an electromagnetic (EM) firehose-like instability, aperiodic and propagating highly obliquely to the magnetic field.
In the present work we investigated this instability under typical conditions found in the heliosphere, calling it the beaming electron firehose instability (BEFI). Thus, we assumed a specific parameterization of the plasma system, including a background embedding plasma of electrons and ions (protons). Counter-beaming electrons penetrating the background solar wind are often reported by in-situ observations, in various contexts such as interplanetary shocks, corotating interaction regions (CIRs), and closed magnetic field topology specific to coronal mass ejections (CMEs).

We relied on such observations to define the plasma model introduced in section 2, and to identify the conditions found favorable to BEFI, see parametric cases in  Table~\ref{t1}. 
In section~\ref{befi} we described the linear spectra of unstable waves for the selected cases in Table~\ref{t1}, varying the relative densities and thermal speed of the electron components. 
Particularly relevant for BEFI, are the regimes identified in Figs.~\ref{f1}-\ref{f5}, for cases 1.a-1.d, 2.a - 2.d and 3.a-3.c, when BEFI is either solely predicted, or has (maximum) growth rates much higher than all the other instabilities in the spectra.
The influence of background population can be quantified in terms of relative density and thermal spread. 
For the cases studied here,  BEFI growth rates are significantly reduced if relative beam densities are less than 20\% of the total density (implying background electrons with relative density exceeding 80\%), making the existence of this instability critical. 
For a slightly cooler background population, compare for instance cases 1 with cases 2, the range of unstable wave-numbers increase. 
Similar effect is obtained in cases 3 for a slightly cooler beam. 
However, for even lower thermal speeds or higher densities of the beams, e.g., in cases 4, the (maximum) growth rates become dominated by the  electrostatic (ES) instabilities at lower angles of propagation,  as already shown in \cite{Lopez-etal-2020b} and \cite{Moya-etal-2022}.

Linear properties of dispersion and stability, including the instability thresholds led us to the conclusion that BEFI is analogous to heat-flux instabilities generated by unidirectional electron strahls/beams in the solar wind. 
BEFI is however triggered by a double heat-flux, that of the counter-beams (or double strahl) of electrons, but for sufficiently low beaming speeds (or associated heat fluxes), in the range of thermal speed of electron beams.
However, in the present analysis with two electron counter-beams and background populations, the configuration of linear spectra of unstable modes becomes much more complicated. 
Besides the ES instabilities (for higher beaming speeds) we have also identified periodic instabilities (with $\omega_r \ne 0$) that do not appear in the absence of the background electron population, when only symmetric counter-beams are present \citep{Lopez-etal-2020b}. 
These unstable wave modes are specific to asymmetric electron beam-plasma configurations, which here result from the combination of each electron beam with the background population. 
More details can be found in a recent parametric analysis of electron heat-flux instabilities in the solar wind conditions \citep{Lopez-etal-2020a}. 
Future works should also investigate more complex plasma systems with asymmetric counter-beams embedded by background electrons, for which we expect BEFI to become a periodic mode as well, see, for instance, the case in Fig.~7 in \cite{Lopez-etal-2020b}. 
In such a case, BEFI will probably blend more easily with other modes and make them difficult to distinguish. 
From the analogy with the heat-flux instabilities, BEFI compares better with the oblique whistlers, which can contribute to the scattering and relaxation of unidirectional strahls in the solar wind \citep{Micera-etal-2020,Cattell-etal-2020}. 

Particle-in-cell (PIC) simulations confirm the results of the linear kinetic theory (section~4), not only for the conditions in which the BEFI is predicted as primary excitation, with major growth rates, but also when it develops as a secondary instability. 
Tested in the PIC simulations were those cases associated with high growth rates, in order to reduce the computational time and obtain results of increased confidence.
The BEFI fluctuations develop (aperiodically) at highly oblique propagation angles to the magnetic field, in agreement with the wavenumber and angular dispersion of the (initial) linear growth rates. 
Moreover, levels reached by these fluctuations are diminished with increasing the presence of background electrons, also contrasting to the results in \cite{Lopez-etal-2020b} obtained in the absence of of background electrons.
In the regimes of competition with ES instabilities, BEFI still develops as a secondary but sufficiently robust instability to produce intense EM fluctuations, long lasting in time up to their saturation. 
Therefore, we can expect BEFI to be involved in the regulation of electron counter-beams with properties similar to those investigated here. 
Our present results should motivate future theoretical and observational studies, to model the evolution of such double electron strahls/beams under the consistent action of BEFI-like fluctuations, and to compare with in-situ observations in space plasmas.

\begin{acknowledgements} 
The authors acknowledge support from the Ruhr-University Bochum and the Katholieke Universiteit Leuven, and Mansoura University. These results were also obtained in the framework of the projects C14/19/089 (C1 project Internal Funds KU Leuven), G.0D07.19N (FWO-Vlaanderen), SIDC Data Exploitation (ESA Prodex-12), Belspo project B2/191/P1/SWiM, and Fondecyt No. 1191351 (ANID, Chile). P.S. Moya is grateful for the support of KU Leuven BOF Network Fellowship NF/19/001 and ANID Chile through FONDECyT grant No. 119135. R.A.L. acknowledges the support of ANID Chile through FONDECyT grant No. 11201048. Powered@NLHPC: This research was partially supported by the supercomputing infrastructure of the NLHPC (ECM-02). We thank the anonymous reviewer for a careful reading of our paper, as well as for the pertinent observations.
\end{acknowledgements}

\appendix

\section{Kinetic dispersion formalism}

Without loss of generality we assume cartesian coordinates ($x,y,z$) with $z$-axis parallel to the magnetic field ${\bf B}$, and with the wave vector ${\bf k}$ in the ($x-z$) plane, such that 
\begin{equation}
  \mathbf{k}\,=\,k_\perp\,\mathbf{\hat{x}}+k_\parallel\,\mathbf{\hat{z}}
\end{equation}
where $\parallel, \perp$ are gyrotropic directions with respect to the magnetic field direction.
From Vlasov-Maxwell equations one can derive the general wave equation \citep{Stix1992}
\begin{equation}
{\bf \Lambda} \cdot {\bf E} = 0 \label{a2}
\end{equation}
and the general dispersion relation for nontrivial (nonzero) plasma modes (far away from the initial perturbation)
\begin{equation}
\Lambda (\omega, k) \equiv {\rm det} [\Lambda_{ij}] = 0, \label{a31}
\end{equation}
in terms of the electric field of the wave fluctuation ${\bf E}({\bf k},\omega)$ and the dispersion tensor ${\bf \Lambda} = |\Lambda _{ij}|$.  
For gyrotropic distribution functions $F_a (v_\perp, v_\parallel)$ of plasma species of sort $a$ (e.g., $a = 0,1,2$ for the electron populations, and $a=p$ for protons) the components of the dispersion tensor read as follows
\begin{eqnarray}
  \Lambda_{ij}(\mathbf{k},\omega)&=&
  \delta_{ij}-\frac{c^2k^2}{\omega^2}\left(\delta_{ij}-\frac{k_ik_j}{k^2}\right)
  \nonumber\\
  &&+\sum_a\frac{\omega_{pa}^2}{\omega^2}\int
  d\mathbf{v}\sum_{n=-\infty}^\infty
  \frac{V_i^nV_{j}^{n*}}{\omega-k_\parallel v_\parallel-n\Omega_a} \nonumber \\
  && \times \left(\frac{\omega-k_\parallel v_\parallel}{v_\perp}\frac{\partial
    F_a}{\partial v_\perp}+k_\parallel\frac{\partial F_a}{\partial
    v_\parallel}\right)+\mathbf{\hat{B}}_i\mathbf{\hat{B}}_j\nonumber\\
  && \times \sum_a\frac{\omega_{pa}^2}{\omega^2}
  \int d\mathbf{v}v_\parallel\left(\frac{\partial F_a}{\partial
    v_\parallel}-\frac{v_\parallel}{v_\perp}\frac{\partial
    F_a}{\partial v_\perp}\right).
    \label{a3}
\end{eqnarray}
with $\mathbf{\hat{B}} = B_0\mathbf{\hat{e}}_3$,
\begin{eqnarray}
  \mathbf{V}^n\, & = & \,v_\perp\frac{nJ_n(b)}{b}\,\mathbf{\hat{e}}_1-iv_\perp
  J_n'(b)\mathbf{\hat{e}}_2+v_\parallel
  J_n(b)\mathbf{\hat{e}}_3, \label{a5}
\end{eqnarray}
$b=k_\perp v_\perp/\Omega_a$, $J_n(b)$ is the Bessel function with $J'_n(b)$ its first derivative, $i$ is the imaginary unit, $c$ is the speed of light, and for each species of sort $a$ $\omega_{pa}=\sqrt{4\pi n_a/m_a}$ is the plasma frequency, $\Omega_a=q_aB_0/(m_ac)$ the gyrofrequency, $q_a$ the charge, $m_a$ the mass, and $n_a$ the number density. 

With the 3-component distribution function in Eqs.~\eqref{e1} and \eqref{e2}, the elements of the dispersion tensor take the following expressions
\begin{eqnarray}
  \Lambda_{11} &=& 1-\frac{c^2k_\parallel^2}{\omega^2}
  +\sum_a\frac{\omega_{pa}^2}{\omega^2}
  \sum_{n=-\infty}^\infty \frac{n^2}{\lambda_a}\Lambda_n(\lambda_a)\mathcal{A}_n \label{a6}
  \,,\\
  \Lambda_{22} &=& 1-\frac{c^2k^2}{\omega^2}
  +\sum_a\frac{\omega_{pa}^2}{\omega^2}
  \sum_{n=-\infty}^\infty \nonumber \\
  && \times \left(\frac{n^2}{\lambda_a}\Lambda_n(\lambda_a)
  -2\lambda_a\Lambda'_n(\lambda_a)\right)\mathcal{A}_n
  \,,\\
  \Lambda_{12} &=& - \Lambda_{21} =  i \sum_a\frac{\omega_{pa}^2}{\omega^2}
  \sum_{n=-\infty}^\infty n\Lambda_n(\lambda_a)\mathcal{A}_n
  \,,\\
  \Lambda_{13} &=& \Lambda_{31} = \frac{c^2k_\perp k_\parallel}{\omega^2}
  +2\sum_a\frac{q_a}{|q_a|}\frac{\omega_{pa}^2}{\omega^2}
  \sqrt{\frac{T_{\parallel a}}{T_{\perp a}}} \nonumber \\   && \times
  \sum_{n=-\infty}^\infty
  \frac{n}{\sqrt{2\lambda_a}}  \Lambda_n(\lambda_a)\mathcal{B}_n
  \,,\\
  \Lambda_{23} &=& - \Lambda_{23} =  -2\sum_a\frac{\omega_{pa}^2}{\omega^2}\frac{|q_a|}{q_a}
  \sqrt{\frac{T_{\parallel a}}{T_{\perp a}}}
  \sqrt{\frac{\lambda_a}{2}} \nonumber \\ && \times
  \sum_{n=-\infty}^\infty
  \Lambda'_n(\lambda_a) \mathcal{B}_n \,,\\
  \Lambda_{33} &=& 1-\frac{c^2k_\perp^2}{\omega^2}
  +2\sum_a\frac{\omega_{pa}^2}{\omega^2}
  \frac{T_{\parallel a}}{T_{\perp a}}
  \frac{U_a}{\alpha_{\parallel a}}
  \left(\frac{U_a}{\alpha_{\parallel a}}+2\xi_a\right) \nonumber \\ && 
  +2\sum_a\frac{\omega_{pa}^2}{\omega^2}
  \frac{T_{\parallel a}}{T_{\perp a}}
  \sum_{n=-\infty}^\infty
  \Lambda_n(\lambda_a)\mathcal{C}_n,
\end{eqnarray}
where $\Lambda_n(x)=I_n(x)e^{-x}$, with $I_n(x)$ the modified Bessel function, and
\begin{eqnarray}  
  \mathcal{A}_n&=&-A_a+(\xi_a-A_a\zeta_a^n)Z(\zeta_a^n)
  \,,\\
  \mathcal{B}_n&=&\xi_a+\left(\zeta_a^n+\frac{U_a}{\alpha_{\parallel a}}\right)
    \mathcal{A}_n
  \\
  \mathcal{C}_n&=&\xi_a\zeta_a^n+\left(\zeta_a^n+\frac{U_a}{\alpha_{\parallel a}}\right)^2 \mathcal{A}_n \,. \label{a14}
\end{eqnarray}
with $Z(x)$ the standard plasma dispersion function for Maxwellian populations \citep{Fried-Conte-1961}. Equivalent expressions for the components of the dielectric tensor are also provided in \citet{Stix1992}, pp. 258--260.

\bibliographystyle{aa}
\bibliography{MS-biblio}
 
\end{document}